\documentclass[reprint,aps,prl,twocolumn,showpacs,preprintnumbers,amsmath,amssymb,amsfonts]{revtex4-1}

\usepackage{graphicx}
 \usepackage{epstopdf}
 \usepackage{color}
\usepackage{siunitx}
\usepackage{natbib}
\newcommand{\degree}{\ensuremath{^\circ} }

\begin{document}

\title{Buffered spectrally-peaked proton beams in the relativistic-transparency regime}

\author{N.~P.~Dover$^{1}$}
\author{M.~J.~V.~Streeter$^{1}$} 
\author{C.~A.~J.~Palmer$^{1}$} 
\author{H.~Ahmed$^{2}$}
\author{B.~Albertazzi$^{3}$}
\author{M.~Borghesi$^{2}$}
\author{D.~C.~Carroll$^{4,5}$}
\author{J.~Fuchs$^{3}$} 
\author{R.~Heathcote$^{5}$}
\author{P.~Hilz$^{6,7}$}
\author{K.~F.~Kakolee$^{2,8}$}
\author{S.~Kar$^{2}$}
\author{R.~Kodama$^{9}$}
\author{A.~Kon$^{9}$}
\author{D.~A.~MacLellan$^{4}$}
\author{P.~McKenna$^{4}$}
\author{S.~R.~Nagel$^{1}$}
\author{D.~Neely$^{4,5}$}
\author{M.~M.~Notley$^{5}$}
\author{M.~Nakatsutsumi$^{3,10}$}
\author{R.~Prasad$^{2}$}
\author{G.~Scott$^{4,5}$}
\author{M.~Tampo$^{9}$}
\author{M.~Zepf$^{2}$}
\author{J.~Schreiber$^{6,7}$}
\author{Z.~Najmudin$^{1}$}

\affiliation{$^{1}$ The John Adams Institute for Accelerator Science, Blackett Laboratory, Imperial College, London SW7 2AZ, United Kingdom}
\affiliation{$^2$ Centre for Plasma Physics, Queen's University Belfast, United Kingdom}
\affiliation{$^{3}$ LULI, \'{E}cole Polytechnique, CNRS, CEA, France}
\affiliation{$^{4}$Department of Physics, SUPA, University of Strathclyde, UK}
\affiliation{$^5$ Central Laser Facility, STFC Rutherford Appleton Laboratory, UK}
\affiliation{$^{6}$ Fakult\"at f\"ur Physik, Ludwig-Maximilians-Universit\"at M\"unchen, D-85748 Garching, Germany}
\affiliation{$^{7}$ Max-Planck-Institut f\"ur Quantenoptik, Hans-Kopfermann-Str. 1, D-85748 Garching, Germany}
\affiliation{$^{8}$ Jagannath University, Dhaka, Bangladesh}
\affiliation{$^{9}$ Graduate School of Engineering, Osaka, Japan}
\affiliation{$^{10}$ European XFEL, GmbH, Albert-Einstein-Ring 19, 22671 Hamburg, Germany}

\date{\today}

\begin{abstract}
Spectrally-peaked proton beams {($E_{p}\approx 8$\,MeV, $\Delta E\approx 4$\,MeV)} have been observed from the interaction of an intense laser ($> 10^{19 }$\,Wcm$^{-2}$) with ultrathin CH foils, as measured by spectrally-resolved full beam profiles. These beams are reproducibly generated for foil thicknesses {5-100\,nm}, and exhibit narrowing divergence with decreasing target thickness down to $\approx 8^\circ$ for 5\,nm. Simulations demonstrate that the narrow energy spread feature is a result of buffered acceleration of protons. Due to their higher charge-to-mass ratio, the protons outrun a carbon plasma driven in the relativistic transparency regime.
 \end{abstract}

\maketitle

The interaction of high intensity lasers with opaque plasma has been widely investigated as a source of multi-MeV ions.  Micron thickness foils irradiated by intense lasers produce sheath fields that accelerate protons to high-energy  \cite{clark2000,snavely2000}.  However, these sheath accelerated beams characteristically have a  thermal spectrum.

The effect of multi-species targets on ion acceleration and spectra has been widely investigated \cite{Decoste1978,allen2003, bychenkov2004,robinson2006}.  Schemes to reduce the energy spread of sheath accelerated beams often rely on spatially localising the protons within a mixed species foil. This was first demonstrated by manufacturing targets with the required ion species localised at the rear of the target \cite{schwoerer2006, hegelich2006}. A similar effect can be achieved by pre-expanding the foil \cite{pfotenhauer2008, dollar2011}, so that the protons can be separated from a trailing lower charge-to-mass ratio host ion species. 

Simulations have shown separation of species can also occur when a single high-intensity pulse interacts with multi-species ultrathin targets \cite{yin2007,qiao2010,yu2010}. Measurements of proton beams accelerated ahead of $\sim$10-100\,\si{\nano \meter} foils have been recently reported \cite{kar2012,steinke2013}. The foils were driven by radiation pressure acceleration (RPA) \cite{zhang2007,robinson2008,klimo2008}, using high-$Z$ foils with $\tau \sim 1$\,ps laser pulses \cite{kar2012} or carbon foils with $\tau \sim 50$\,fs \cite{steinke2013}.  The proton layer was `buffered'  from direct interaction with the laser by the presence of the lower charge-to-mass ratio species, giving a compression in spectrum as well as a modest energy boost. Buffering {may} also insulate protons from transverse instabilities that occur during RPA \cite{palmer2012}.  

Recent experiments have also demonstrated enhanced acceleration from ultrathin targets when the plasma electron density, $n_e$, drops below the relativistic critical density, $\gamma n_c$ \cite{yin2006,yin2011, willingale2009, henig2009,jung2013b}.
Removing protons prior to interaction resulted in higher carbon energies \cite{jung2013}. However, in a multi-species target, protons should exhibit buffered acceleration in this regime also.

We present the first observation of buffered acceleration of protons from ultrathin (5-100 nm) carbon foils in the relativistic transparency regime.  Narrow energy spread proton {beams} were observed using beam profile stack detectors and high resolution spectrometers. This allowed, for the first time, full spatial characterisation of the spectrally modulated beams.  The beams had typical peak energy $E_{p}\approx 8$\,MeV, corresponding to a velocity equal to that of the carbon ion front, and $\Delta E \approx 4$\,MeV.  The full-angle beam divergence reduced with decreasing target thickness down to $\approx 8^\circ$ for 5\,nm foils. Simulations demonstrate that the narrow divergence, narrow energy spread feature is accelerated ahead of, and buffered from, filamentation due to transverse laser-matter instabilities.

The experiment was performed at the Vulcan Petawatt facility at the Rutherford Appleton Laboratory. A single plasma mirror, irradiated at $I_{pm} = 3 \times 10^{14}$\,Wcm$^{-2}$, enhanced the laser contrast to $\approx 10^{-10}$ \cite{thaury2007}.  The resulting ($130\pm20$)\,J, $\tau \approx$\,700\,fs pulse was $f/3$ focussed with $35$\,\% of the energy within a focal spot 1/e$^2$ width $w_{0} = 8$\,\si{\micro \meter} measured at low power. Since this is not measured at high power, this implies a maximum possible vacuum intensity of $I_L = (9 { \pm 2}) \times 10^{19}$\,Wcm$^{-2}$. The laser was either linearly (LP) or circularly (CP) polarised, and normally incident onto diamond-like carbon (DLC) foils of thickness $d \in 5$-$500$\,nm. These comprise $\approx 90$\%\,C/10\%\,H by number\cite{palmer2012}, and additionally feature a nm scale hydrocarbon impurity layer. 

\begin{figure}  
 \centering
 \includegraphics[width=0.45\textwidth]{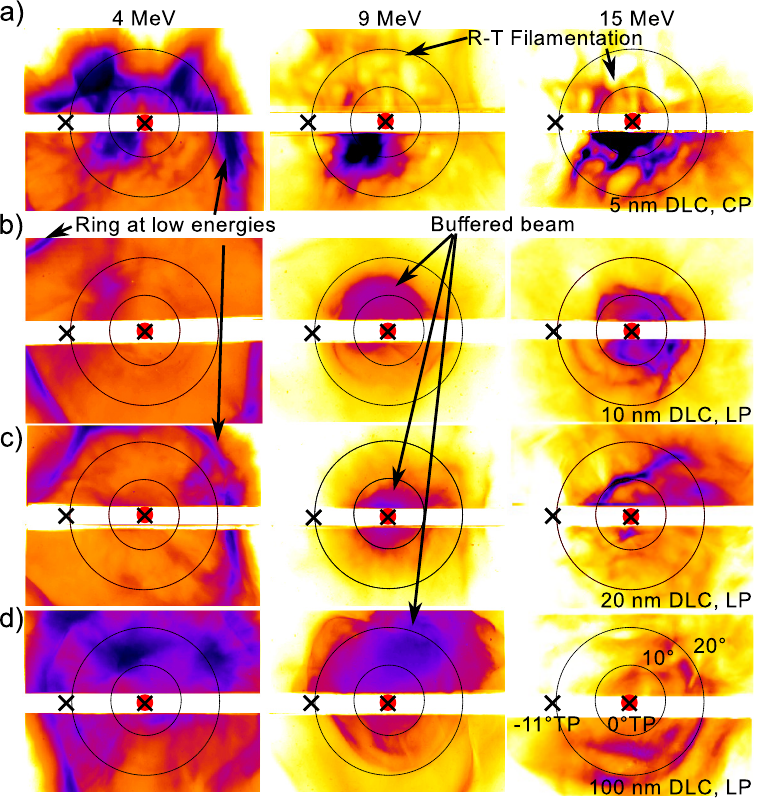} 
 \caption{a-d) Proton beam profiles from 5, 10, 20 and 100\,nm target using CP, LP, LP and LP respectively at $4, 9, 15$ MeV. Red dot represents laser-axis, and black rings are at 10, 20$^\circ$ full-cone angle.  The directions of 0\degree and -11$^\circ$\,TP are indicated with black crosses. (20$^\circ$\,TP is off the right edge of RCF).  All films are contrast enhanced to see detail, and dark colour represents higher dose.}
\label{profiles}
 \end{figure}

The proton beam profile was diagnosed using a radiochromic film (RCF) stack 10.6\,cm from the interaction, allowing observation within  a $\lesssim 25^\circ$ full angle. Due to the enhanced stopping of protons at the end of their range, the dose deposited in each layer of the stack is dominated by a small range of proton energies, $E$. Carbon ions would require $> 80$\,MeV to pass through to the first RCF. The small numbers of carbon ions measured at these high energies means their contribution to the dose could be neglected.  Higher energy resolution spectra were recorded by three Thomson parabola (TP) spectrometers. The TP sampled the beam at 0$^\circ$, -11$^\circ$ and +20$^{\circ}$ behind a horizontal gap in the stack (see fig.~\ref{profiles}). The TP use co-linear magnetic and electric fields to separate ion species by charge-to-mass ratio and disperse them by energy before they are detected by calibrated Fujifilm imaging plate \cite{calib}.

Examples of proton beam profiles are shown in fig.~\ref{profiles} for both CP and LP with targets of $d = 5$-100 nm. The gap for TP access is shown to scale. For protons with energy $E = 4$ MeV, the beam is dominated by an annular ring with divergence angle $>20^{\circ}$. This ring structure was characteristic for all polarisations and thicknesses. On occasion the ring structure was directed into the -11$^{\circ}$\,TP revealing a high-flux 
($\frac{{\rm d}^{2}N}{{\rm d}E {\rm d}\Omega} \sim 1$-$5\times10^{12}\,{\rm MeV^{-1}sr^{-1}}$) broadband beam with energy up to $E_{\text{max}}\approx 5$ MeV. 

\begin{figure}  
 \includegraphics[width=0.47\textwidth]{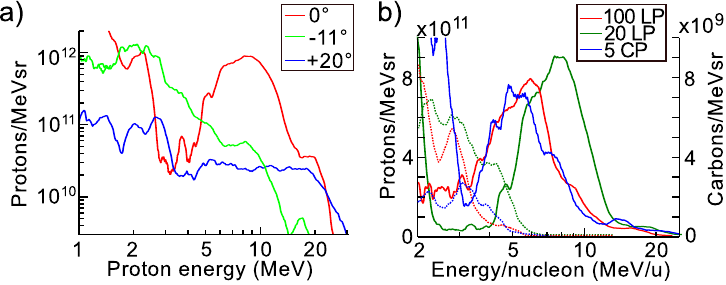}
 \caption{a) H$^{+}$ spectra from all TPs from LP, 20\,nm target shot shown in fig.~\ref{profiles}c.   b) H$^{+}$ (solid) and C$^{6+}$ ion (dashed) spectra for three different thicknesses (in nm).}
\label{spectra}
 \end{figure}

On the 9\,MeV profiles the annular ring has disappeared for all target thicknesses.
Instead, there is a narrower divergence circular beam near the laser-axis for all targets between 5 and 100\,nm. The central beam has a full angle divergence $\theta<20^{\circ}$, and becomes less divergent for thinner targets. For $d \leq 20$ nm, there is a filamented halo surrounding the central beam, similar to previous reports of a Rayleigh-Taylor-like instability \cite{palmer2012}.  The filamentation is not transposed onto the central beam, suggesting the source of the two populations is different.  By the $E= 15$\,MeV slice, the distinct central beam has disappeared in all but the thinnest targets.  

Proton spectra from all three TPs for a 20 nm target are given on a log scale in fig.~\ref{spectra}a, showing a pronounced peak with energy $E_{p}=8$\,MeV, and energy spread $\Delta E \approx 4$\,MeV, but only at 0$^\circ$. The peaked spectrum is shown on a linear scale in fig.~\ref{spectra}b, along with on-axis spectra for $d=5, 100$\,nm targets. Using the divergence from the 9\,MeV beam profile, integrated over the spectrum gives $(3\pm1)\times 10^{11}$ protons within the spectral FWHM for $d= 20$~nm, with a conversion efficiency of laser energy of $\approx  0.25\%$.
 The corresponding broadband carbon spectrum are also plotted, and  show a maximum velocity $\approx 3\times10^7\,{\rm ms}^{-1}$ that corresponds to the start of the H$^{+}$ peak. The protons in the peak have evidently outrun the C$^{6+}$ front, and are \emph{shielded} from the laser-plasma instabilities driven in the denser carbon plasma.

On shots featuring this buffered beam, a highly divergent but comparatively low flux proton beam was also measured on both the RCF stack and the TP reaching higher energies ($E_{\text{max}}\approx20$-30\,MeV). This component is visible at $\approx15$\,MeV in fig.~\ref{profiles}a-d by enhancing the picture contrast.  This population of protons was sufficiently divergent to be routinely observed on the 20$^\circ$\,TP, well outside of the divergence of the ring structure.  

We note that these beam profile measurements highlight the need for both beam profile and spectral measurements when reporting narrow energy spread ion beams from laser matter interaction, as the beam properties can vary strongly with angle and cannot be extrapolated by sampling over limited angular acceptance.
 
\begin{figure}[t]
 \includegraphics[width=0.47\textwidth]{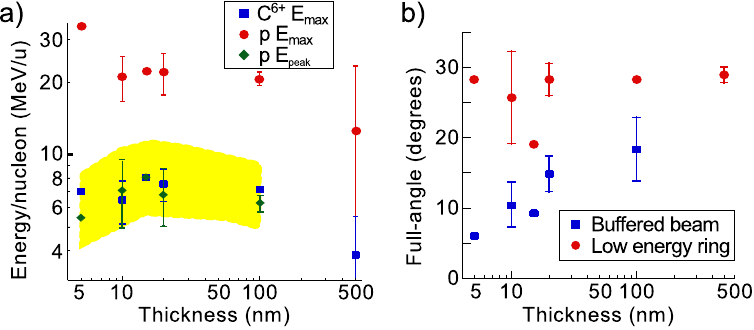}
 \caption{As a function of $d$: a) max energy per nucleon for H$^{+}$ (red circles) and C$^{6+}$ (blue squares) and energy of H$^{+}$ peak (green diamonds) with FWHM energy spread shaded yellow;  b) divergence $\theta$ of central beam (blue squares, at $E = 9$\,MeV) and low energy ring (red circles, at $E = 4$\,MeV).  Error bars are standard deviation of multiple shots.}
\label{fig3}
 \end{figure}

The buffered on-axis beam was generated reproducibly on all shots for foil thicknesses between 10 and 100\,nm for both CP and LP.  The energy of the peak and its energy spread is shown over this range in fig.~\ref{fig3}a, including both LP and CP.  Both polarisations showed similar results for $d\in 20$-$100$ nm. However for $d=5$\,nm, only CP resulted in spectrally peaked beams, LP resulted in a thermal spectrum with $E_{\text{max}}\approx 5$\,MeV. Due to this marked difference this data has not been included in fig.~\ref{fig3}. CP allows acceleration of thinner targets, presumably due to reduced target heating resulting in the target remaining intact for longer.  No peak was observed for $d = 500$\,nm.

Also plotted in fig.~\ref{fig3}a are the maximum carbon and proton energy per nucleon, $E_{\text{max}}$, as observed on any of the TP, averaged over multiple shots.  The maximum energy was most often off-axis on the 20$^{\circ}$\,TP.  $E_{\text{max}}$ for the C$^{6+}$ ions correlates well with the proton peak energy, $E_{p}$, confirming that their velocities are linked. Fig.~\ref{fig3}b shows that the divergence of the central beam {\emph{decreases}} with decreasing target thickness, as was apparent in fig.~\ref{profiles}, down to $\approx 8^{\circ}$ for $d=5$ nm on the 9\,MeV slice. By contrast, the low energy ring size, measured at 4\,MeV, remains around $ 26\pm 5 ^{\circ}$, indicating that these are two separate populations.   

The experiment was simulated with the 2D particle-in-cell (PIC) code OSIRIS \cite{fonseca2002} in a box of $160\times80$\,\si{\micro \meter} with cell size $4\times10$\,nm.  The target was a  plasma composed of 90/10\%\ of C$^{6+}$/H$^{+}$ by charge density with $n_{e} = 1000\,n_c$, and 2500 particles per cell for each species. Foil thickness, $d$, was varied between 5 and 100\,nm.  The electron temperature was initialised at $T_{e} = 200$\,eV, resulting in some expansion of the target before the arrival of the high intensity laser pulse, relaxing the resolution constraints for the simulation \cite{yin2011}.  The laser was initialised with CP and gaussian transverse and longitudinal field profile with focal spot size $w_{0} = 8\,\si{\micro \meter}$, pulse length $\tau=650$ fs, and $\displaystyle a_0\equiv {eE_{0}}/{mc\omega_{0}} = 3$.  

 \begin{figure}  
 \includegraphics[width=0.47\textwidth]{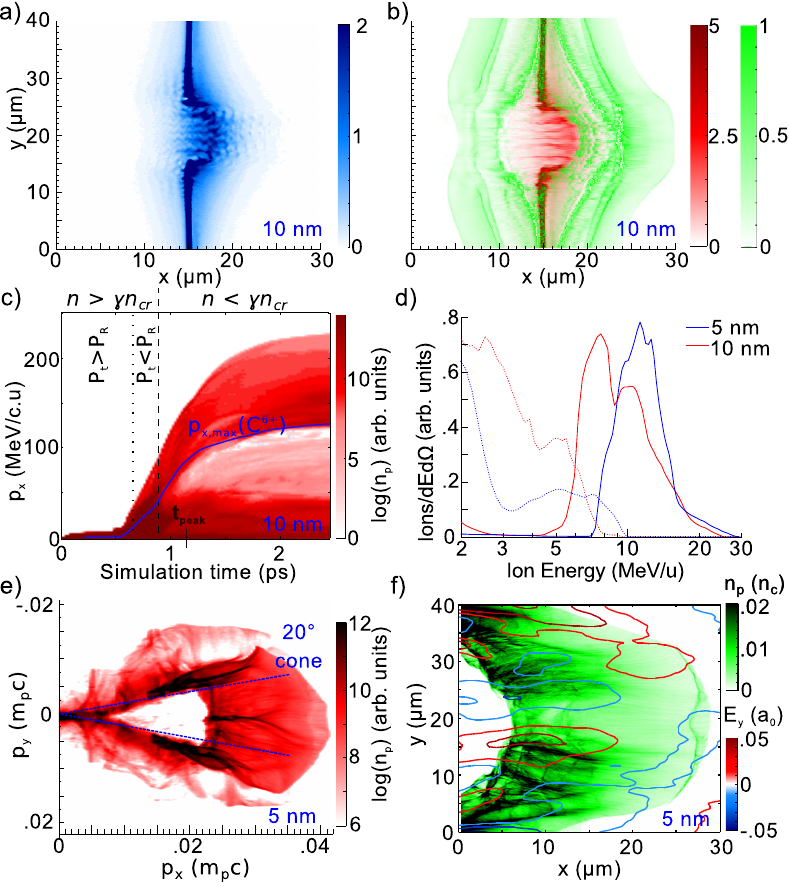}
 \caption{From simulation: charge density of  a) electrons and b) C$^{6+}$ (red) and H$^{+}$ (green) near peak of pulse, $t_{\text {peak}}$, for $d=10$\,nm;  c) Evolution of $p_x$ for H$^{+}$ on-axis, with $p_{x,\text{max}}$ for C overlaid (blue line) for $d=10$\,nm; vertical dashed (dotted) lines indicate different density (pressure) regimes; d) spectra of H$^{+}$ witnessed in $\approx 1$\,msr on-axis for $d=5, 10$\,nm;  e) $p_xp_y$ of H$^{+}$ at end of acceleration for $d=5$\,nm; f) $n_{p}$ near end of pulse ($t_{\text {peak}}+600$\,fs) with overlaid $E_y$ contours for $d=5$\,nm.
 }
\label{fig4}
 \end{figure}
For a foil of $d=10$\,nm, the thickness is comparable to the skin depth $c/\omega_p$. Nearly all the target electrons in the focal spot are heated to the instantaneous ponderomotive potential and sheath formation is enhanced, causing the target to expand during the rising edge of the laser. As the laser intensity increases, the radiation pressure, $P_R$, of the pulse exceeds the plasma thermal pressure $P_t$, or $I_L/c>n_ek_bT_e$, and the front surface begins to recede. In this stage, the front surface is prone to transverse Rayleigh-Taylor like instability\cite{palmer2012}, enhancing decompression of the target and quickly causing it to become relativistically transparent to the laser.  The electron and ion charge density in this phase is shown in fig.~\ref{fig4}a,b.  There is a longitudinal electric field $E_x$ inside the transparent region.  Due to their higher charge-to-mass ratio, the protons accelerate faster than the C$^{6+}$ ions, and therefore quickly spatially separate (fig.~\ref{fig4}b). 

Fig.~\ref{fig4}c shows the temporal evolution of the on-axis forward momentum $p_x$ of the protons along with $p_{x,\text{max}}$ of the carbon ions. Once the target becomes relativistically underdense, as indicated by the dotted black vertical line, both the electron temperature and the ion kinetic energy quickly increase. The electrons are volumetrically heated and exchange energy with the ion species via streaming instabilities \cite{yin2006,yin2007}. The ions energy gain occurs mostly in this phase where the target is relativistically transparent ($n_{e}<\gamma n_{cr}$), and is not due to radiation pressure \cite{yin2011}.  Energy is gained from the fields within the target due to the volumetric heating \cite{yin2006}, and also those due to the sheath fields ahead of the carbon front.

Below a cutoff energy $E_c$, there are almost no protons, which results in peaked spectra when sampled on-axis similar to those observed experimentally (fig.~\ref{fig4}d). Due to their lower charge-to-mass ratio, the protons quickly outrun the carbon ions, gaining a minimum velocity equal to the velocity at the carbon ion front $v_{c}$. 
In this case, this happens shortly after the target becomes relativistically transparent, but in general depends on the ratio of protons to carbon ions. Though the carbon ions have a strong transverse modulation, which would be witnessed as a strongly filamented carbon ion beam, most of the protons have been buffered from this filamentation.

At later times, the minimum proton energy in the buffered layer increases, following the energy of the carbon ion front, whose energy rises quickly initially before saturating. The maximum proton energy increases due to acceleration in the sheath. This leads to a high energy proton component with wide angular divergence, as witnessed as a low signal background in the experiment.

Expulsion of protons from the carbon sheath is apparent in their $p_xp_y$ phase space as seen for the thinnest target, $d=5$\,nm, in fig.~\ref{fig4}e. An additional low divergence component ($\approx\,4^\circ$) is also visible close to the axis.  In the initial phase of the expansion, electron heating is localised to a small transverse extent similar in size to the longitudinal extent of the expanding plasma $\sim w_{0}$.  The appreciable transverse to longitudinal electric field ratio $E_y/E_x \approx \frac{\partial{n_e}}{\partial{y}}/\frac{\partial{n_e}}{\partial{x}} \sim \frac{1}{3}$, results in a rapid transverse expansion and a proton density minima on laser-axis, shown in fig.~\ref{fig4}f.   The remnants of this initial expansion are responsible for the ring observed at low energy on our RCF.  This is a feature of longer pulses ($\tau \approx 1$\,ps) and is in contrast to the case for ultrashort pulses \cite{bin2013}. 

The sheath of this self-generated cone has a focussing effect on trailing protons still being accelerated within the evacuated region. This is similar in action to other laser-triggered charged particle lenses \cite{Toncian2006, Kar2008}. The focussing field, shown by the contours in fig.~\ref{fig4}f, is maintained until the end of the interaction, producing a collimated beam on-axis. For thicker targets, there are sufficient protons in the target that a larger fraction remain in the central region, reducing the collimating $E_y$ field and leading to a wider divergence for the buffered protons.  LP simulations showed very similar behaviour to CP, apart from a slightly faster initial expansion.

Further simulations were performed for $d=20$\,nm targets, varying $a_0$. For $a_0 < 2$, the target remains overdense throughout the interaction, and the radiation pressure never overcomes the plasma pressure, leading to sheath acceleration being dominant. For $a_0 > 4$, the target becomes transparent before the peak of the pulse, and both species gain energy during the relativistic transparency stage, as in fig.~\ref{fig4}c.  However, for $a_0 = 8.5$, as expected for the experimental parameters, $E_{p} = 35$\,MeV, significantly higher than experimentally measured.  The lower energies observed imply a lower intensity at focus, which is likely to be due to poor near-field uniformity of the laser combined with hydrodynamic expansion of the plasma mirror for a $\tau \approx 1 $\,ps pulse. 
For  $d=20$\,nm, the simulations suggest $E_{p} \propto a_0$, following a ponderomotive scaling, although energies can be optimised for a target thickness dependent on the laser intensity \cite{yin2006}.  The peak proton energy is also strongly dependent on the ratio of carbon to proton mass in the target \cite{jung2013}, with a trade-off between higher energy but fewer protons.

In conclusion, by using relatively long pulse and ultrathin carbon foils we can simultaneously operate in the relativistic transparency regime with its enhanced efficiency and generate buffered, self-collimated beams with reduced energy spread. The buffering is of importance for the acceleration of ultrathin foils that are naturally prone to instabilities, making this an interesting source of laser generated protons. 

\begin{acknowledgments}
The acknowledge funding by EPSRC/RCUK grants EP/E035728/1, EP/K022415/1 and STFC grant ST/J002062/1. We thank the {\sc Osiris} consortium (UCLA/IST) for use of {\sc Osiris}, and the support of the LMU's MAP-service centre to provide DLC foils.
\end{acknowledgments}

\end{document}